\documentclass[12pt]{book}

\usepackage[dvips]{graphicx,color}
\usepackage{makeidx,physics,cosmology}

\makeauthorindex
\makeindex

\BookTitle{Frontier in Astroparticle Physics and Cosmology}
\CopyRight{\copyright 2004 by Universal Academy Press, Inc.}

\begin{document}

\BookTitle{\itshape Frontier in Astroparticle Physics and Cosmology}
\CopyRight{\copyright 2004 by Universal Academy Press, Inc.}
\pagenumbering{arabic}

\chapter{Cosmology of Brane-worlds}

\author{%
David LANGLOIS\\
{\it  GReCO, Institut d'Astrophysique de Paris (CNRS), \\
98bis Boulevard Arago, 75014 Paris, France}}
%
%
\AuthorContents{D. Langlois} 

\AuthorIndex{Langlois}{D.} 

\section*{Abstract}
This talk presents an overview of  the  brane cosmology scenario, based 
on the idea that our Universe is a 3-brane embedded in a five-dimensional 
anti-de Sitter bulk space-time. Special emphasis is put on the novel 
features of this scenario: an unconventional cosmological evolution 
at high energy densities, i.e. in the early universe, and dark radiation, 
that embodies the gravitational effects of the bulk onto the brane, and 
which is shown to be generated during the high energy era by the production
of bulk gravitons.

\section{Introduction} 
In this contribution, I review the basic ingredients of the so-called 
Randall-Sundrum brane cosmology. This new cosmological scenario
is based on the  assumption that our universe is  
a {\it brane}: a sub-space embedded in a  bulk spacetime, 
with a single extra  dimension. 
In contrast with other braneworld models, the 
self-gravity of the brane is  taken into account. 

As will be recalled in this contribution, the cosmology for such 
a brane-universe is modified in two respects: 
\begin{itemize}
\item  the 
Friedmann equation is modified at high energy;
\item the bulk influences the cosmological evolution 
via an addititional term, usually called  dark radiation or Weyl radiation.
\end{itemize}
Whereas the first modification has an impact only during the very 
early universe, since it is significant only at high energy, the 
second effect could have observable consequences today, as discussed below.

In this contribution, I do not discuss other important topics in the context 
of brane cosmology, such as the issue of cosmological perturbations, 
since this topic will be discussed by Roy Maartens. Finally, for 
the reader who wants to learn more on this subject, he/she can find in 
the literature 
several detailed reviews \cite{reviews} 
that cover much more  than the present contribution.

\section{Homogeneous brane cosmology}

\subsection{The model}
As in standard cosmology, the starting point is to 
 assume {\it homogeneity}
 and {\it isotropy} in the three ordinary spatial dimensions. However, in 
the context of brane cosmology, these symmetries cannot be extended to the 
extra dimension, since the presence of the brane itself 
breaks homogeneity along the extra dimension.
As a consequence, all physical quantities depend  on time and on 
the extra dimension.

In a suitable coordinate system (Gaussian Normal), 
the five-dimensional metric can be written in the form 
\begin{equation}
ds^2=- n^2(t,y)dt^2+a^2(t,y)d{\vec x}^2+dy^2,
\end{equation}
with the brane located at $y=0$. To obtain the equations governing the 
cosmological evolution, one substitutes this ansatz into
 the five-dimensional Einstein equations (with the bulk cosmological 
constant $\Lambda$)
\begin{equation}
G_{AB}+\Lambda g_{AB}=\kappa^2 T_{AB}
\end{equation}
where  the energy-momentum tensor, assuming a  bulk otherwise empty, 
is due to the brane matter and  thus given by
\begin{equation}
T_A^B=Diag(-\rho_b(t), P_b(t), P_b(t), P_b(t), 0)\delta(y).
\end{equation}

\subsection{The cosmological evolution}
It turns out that it  possible to solve explicitly 
the five-dimensional Einstein's 
equations (see \cite{bdel99}). Specializing the obtained 
solution at the brane location (denoted by the subscript 'b'), 
one  finds   the {\it modified 
Friedmann equation} \cite{bdl99,cosmors,bdel99}:
\begin{equation}
\label{fried}
H_b^2\equiv {\dot a_b^2\over a_b^2}={\Lambda\over 6}+{\kappa^4\over 36}
\rho_b^2+{{\cal C}\over a_b^4}
\end{equation} 
where ${\cal C}$ is an integration constant.
It can also be shown that, for an empty bulk, the  
usual  conservation equation  holds, which implies
\begin{equation}
\dot\rho_b+3H_b(\rho_b+P_b)=0.
\end{equation}

 For $\Lambda=0$ and ${\cal C}=0$,  the bulk is 5-D Minkowski and  the cosmology 
is highly unconventional since the Hubble parameter is proportional to 
the brane energy density \cite{bdl99}. This has the 
unfortunate   consequence  to 
ruin the standard nucleosynthesis scenario that relies on the evolution of
the expansion rate with respect to the relevant microphysical 
interaction rates. 

To obtain a viable brane cosmology scenario, the simplest way is to 
use the idea of  Randall and Sundrum \cite{rs99b}, 
i.e. to make the following 
assumptions:
\begin{itemize}
\item Consider a bulk with a negative cosmological constant $\Lambda<0$
 \item 
 Assume the brane is endowed with an intrinsic tension $\sigma$, so that 
$\rho_b(t)=\sigma+\rho(t)$,
where  $\rho(t)$ is the energy density of usual cosmological matter.
\end{itemize}
With these assumptions, the Friedmann equation  (\ref{fried}) yields
\begin{equation}
H_b^2=\left({\Lambda\over 6}+{\kappa^4\over 36}\sigma^2 \right)
+{{\kappa^4\over 18}\sigma} \rho +{\kappa^4\over 36}\rho^2
+{{\cal C}\over a_b^4}.
\end{equation}
One recovers   {\it approximatively}  the  usual 
Friedmann  equation if 
\begin{equation}
{\Lambda\over 6}+{\kappa^4\over 36}\sigma^2= 0,
\end{equation}
which is the condition imposed by Randall and Sundrum in their 
(non-cosmological) model to recover standard gravity, and which also implies
 \begin{equation}
\label{G}
8\pi G\equiv {\kappa^4\over 6}\sigma.
\end{equation}
However, this Friedmann equation is characterized by two new features:
\begin{itemize}
\item A $\rho^2$ term, which dominates at high
energy;
\item A radiation-like term,  ${{\cal C}/ a_b^4}$, usually called 
{\it dark radiation}.
\end{itemize}
The cosmological evolution undergoes  a transition from a high 
energy regime, $\rho\gg \sigma$, characterized by 
an unconventional behaviour of the scale factor, into a 
low energy regime which reproduces our standard cosmology. For 
${\cal C}=0$ and an equation of state $w=P/\rho=const$, one can solve 
analytically the evolution equations and one finds
\begin{equation}
a(t)\propto t^{1/q}\left(1+{q\, t\over 2\ell}\right)^{1/q},
\qquad q=3(1+w).
\end{equation}
One clearly sees the transition, at the epoch $t\sim \ell$, between 
the early, unconventional, evolution $a\sim t^{1/q}$ and the standard
evolution $a\sim t^{2/q}$.

\subsection{The bulk point of view}
The above cosmological evolution can be obtained  from a very different 
perspective \cite{kraus}, 
by starting from a {\it static} bulk metric, which, because of 
the cosmological symmetries and the (negative) cosmological constant, must be 
 AdS-Schwarzschild in five dimensions:
\begin{equation}
\label{ads}
ds^2=-f(R)dT^2+{dR^2\over f(R)}+R^2 d\Sigma_k^2,
\quad
f(R)=k+{R^2\over \ell^2}-{{\cal C}\over R^2}, \quad k=0,\pm 1.
\end{equation}
In this coordinate system, the brane is {\it moving} and the so-called 
junction conditions 
$[K_{\mu\nu}]=-\kappa^2\left(S_{\mu\nu}-(S/3)g_{\mu\nu}\right)$
give the modified Friedmann equation obtained above.

\subsection{Constraints on the parameters}

This cosmological scenario is essentially characterized by the value of 
the ``fundamental mass scale'' $M_5$ defined as
 $\kappa^2= M_5^{-3}$, since the other parameter, the AdS lengthscale 
$\ell\equiv \sqrt{-\Lambda/6}$, is related to $M_5$ via the relation 
(\ref{G}), i.e.
\begin{equation}
M_{Pl}^2=M_5^3\ell,
\end{equation}
which defines the four-dimensional Planck mass in this set-up. 
The scenario must satisfy two constraints:
\begin{itemize}
\item be compatible with the nucleosynthesis scenario, which means that 
the high energy regime, mentioned above, must take place before 
nucleosynthesis. This requires $\sigma^{1/4} > 1$ MeV, and since 
$\sigma= 6/(\kappa^2\ell)=6 M_5^6/M_P^2$, this gives the constraint
$M_5> 10^4$ GeV.

\item be compatible with the gravity experiments on small scales, 
which presently require $\ell < 0.1$ mm. This implies $M_5 > 10^8$ GeV.

\end{itemize}
As will be detailed in the next section, another observational constraint
applies to the dark radiation constant ${\cal C}$.

\section{Dark radiation and the production of bulk gravitons}

So far, the bulk has been assumed to be {\it strictly empty}, apart from 
the presence of the brane.
However, the fluctuations of brane matter generate bulk 
gravitational waves. Equivalently, at smaller scales, the scattering of brane 
particles can produce bulk gravitons: 
$$
\psi+{\bar \psi}\rightarrow G
$$
Therefore, the fact that the homogeneity hypothesis in the brane 
is not exactly satisfied necessarily leads 
to the presence of a flow of gravitons
in the bulk.

\subsection{Emission by the brane}
The production of  
gravitons results into an energy loss for ordinary matter, which can be 
 expressed as  
\begin{equation} 
{d\rho \over dt}+ 3H(\rho+P)=- 
\int\frac{d^3p}{\left(2 \pi\right)^3} 
{\bf C}\left[f\right],  
\label{prl16} 
\end{equation} 
with  the collision term
\begin{equation} 
{\bf C}\left[f\right]={1\over 2} 
\int \frac{d^3p_1}{\left(2 \pi\right)^3\,2E_1}\,  
\frac{d^3p_2}{\left(2 \pi\right)^3\,2E_2} 
\,\sum \left\vert {\cal {M}}\right\vert^2\,f_1\,f_2\, 
\left(2\pi\right)^4\,\delta^{(4)}\left(p_1+p_2-p\right)\,\, , 
\label{prl17} 
\end{equation} 
where ${\cal {M}}$ is the scattering amplitude for the process in  
consideration (the indices $1$ and $2$ correspond to the scattering 
particles $\psi$ and $\bar\psi$). The summed squared amplitude  is given by 
\begin{equation} 
\sum \left\vert \cal{M}\right\vert^2=\hat g\,\frac{\kappa^2}{8\,\pi}\,s^2, 
 \qquad \hat g=(2/3)g_s+4g_v +g_f
\end{equation} 
with $s=\left(p_1 + p_2 \right)^2$.
When the brane matter is in thermal equilibrium with a temperature 
$T$, one finds 
\begin{equation}
\label{emission}
\dot\rho+4H\rho=-{315\, \over 512\, \pi^3}
\, \hat g \, \kappa^2\,  T^8.
\end{equation}

\subsection{Bulk description of the radiation brane}
 It is much more difficult to describe what happens in the bulk. 
A possibility is to model 
the system Bulk + Radiating Brane by  a generalized 
five-dimensional Vaidya metric  \cite{lsr02}
\begin{equation}
ds^2=-f(R,v)dv^2+2dR dv +R^2 d{\bf x}^2, \quad 
f(R,v)=\mu^2 R^2-{{\cal C}(v)\over R^2}.
\end{equation}
describing an {\it ingoing} radiation flow.
Here, $v$ is a null coordinate. 
If ${\cal C}(v)$ is constant, one 
recovers the AdS-Schwarzschild metric (\ref{ads}).
The generalized Vaidya's metric is a solution of the five-dimensional 
Einstein's equations with
\begin{equation}
T_{AB}={\cal F} \, k_A k_B,  \qquad k_Ak^A=0, 
\end{equation}
which means that the gravitons must be radial. 

This description is in  general too restrictive. More generally, the bulk 
must be seen as  filled with a gas of gravitons with a distribution 
function $f$. Their energy-momentum
tensor is given by
\begin{equation} 
{\cal T}_{AB}=\int d^5p \ \delta\left(p_Mp^M\right)\sqrt{-g}\,f\, p_Ap_B, 
\label{T_integ} 
\end{equation}  
 From the 5D Einstein equations,
one can derive effective 4D Einstein equations \cite{sms99}, 
which in the homogeneous 
case yield 
\begin{itemize}
\item the  Friedmann equation
\begin{equation}
H^2 =   
{{\kappa _4^2} \over 3}\left[  
\left(1+{\rho\over 2\sigma}\right)\rho+ \rho_{\rm{D}}  \right],  
\label{Hubble} 
\end{equation} 
\item the non-conservation equation for brane matter
\begin{equation} 
\dot{\rho}+3\,H\,\left(\rho+p\right)=2\,{\cal {T}}_{RS}\,n^R\,u^S\,\,
\end{equation} 
which must be identified with (\ref{emission});
\item the non-conservation equation for the ``dark radiation'' energy density
$\rho_{\rm{D}}$ (which includes all effective contributions from the bulk)
:
\begin{equation}
 \dot \rho_{\rm{D}}+4H\rho_{\rm{D}} 
  =-2\left(1+{\rho\over\sigma}\right){{\cal T}}_{AB} u^A n^B   
  -2 H\ell \, {{\cal T}}_{AB} n^A n^B\,. 
\end{equation} 
\end{itemize}
On the right hand side of this last equation, we find two terms with 
opposite signs: the first term, due to the energy flux from the brane 
into the bulk, contributes positively and thus increases the amount of 
dark radiation whereas the second term, due to the pressure along 
the fifth dimension, decreases the amount of dark radiation.
To estimate quantitatively these terms, one needs to determine the 
distribution of the bulk gravitons. 

\begin{figure}[t]
  \begin{center}
    \includegraphics[height=13pc]{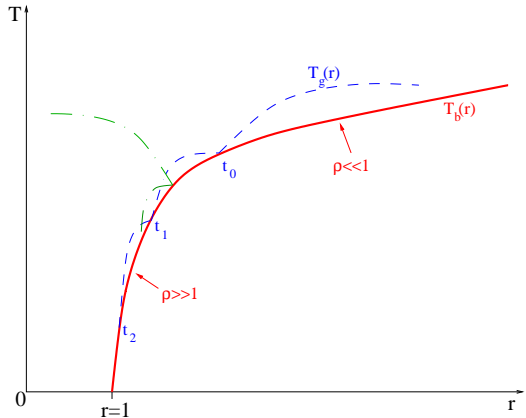}
  \end{center}
  \caption{The asymmetry $A_2$ as a function of $x$.}
\end{figure}

This task was undertaken recently \cite{ls03} by assuming that the 
bulk background stays approximately AdS during the whole evolution. For this
background, one can compute analytically the trajectories of the bulk 
gravitons. Figure 1 shows such a few trajectories, together with the 
trajectory of the brane endowed with relativistic matter. A peculiar feature 
is that many (non-radial) gravitons  tend to come back onto 
the brane and  bounce off it. This gives a significant contribution 
to the 
transverse pressure effect, which almost, although not quite, compensates
the flux effect. 

The amount of dark radiation produced, in terms of the ratio 
$\epsilon_D\equiv \rho_D/\rho$ is plotted on Fig. 2 as a function 
of the initial radiation energy density (and compared with the analytical 
results of \cite{hm01} and \cite{lsr02}).
\begin{figure}[t]
  \begin{center}
    \includegraphics[height=20pc,angle=-90]{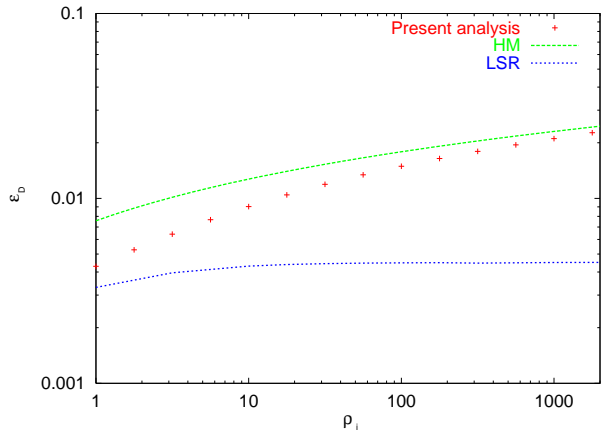}
  \end{center}
  \caption{Amount of dark radiation $\epsilon_D=\rho_D/\rho$ as 
a function of the initial energy density on the brane 
$\rho_i$ (in units of $\sigma^4$), 
computed numerically in \cite{ls03} and compared with HM \cite{hm01}
and LSR \cite{lsr02}.}
\end{figure}

\subsection{Observational constraints}
The computed amount of dark radiation can be confronted to observations 
(see e.g. \cite{Ichiki:2002eh}). 
Indeed, since 
dark radiation behaves as radiation, it must satisfy the  
 nucleosynthesis  constraint on the number of {\it additional 
relativistic degrees of freedom}, usually expressed in terms 
of the extra number of light neutrinos $\Delta N_\nu$.
 The relation between $\Delta N_\nu$ and $\epsilon_D$ is given by
\begin{equation}
\epsilon_D={7\over 43}\left({g_*\over g_*^{\rm nucl}}\right)^{1/3}
\Delta N_\nu,
\end{equation}
where $g_*^{\rm nucl}=10.75$ is the number of degrees of freedom at
nucleosynthesis (in fact before the electron-positron annihilation).
Assuming $g_*=106.75$ (standard model), this gives
$\epsilon_D\simeq 0.35 \Delta N_\nu$. 
The typical constraint from nucleosynthesis 
\begin{equation}
\Delta N_\nu < 0.2
\end{equation}
implies
\begin{equation} 
\epsilon_D\equiv {\rho_D\over\rho_r} < 0.03 \left({g_*\over 
g_{*}^{\rm nucl}}
\right)^{1/3}
\end{equation}
which gives $\epsilon_D < 0.07$ with the degrees of freedom 
 of the standard model.

 \section{Conclusions}
The various models of extra dimensions with branes have raised a considerable 
interest in the last few years, motivated by their more or less direct 
connections with the recent developments in string/M theory. To make 
further progress in this direction, it is important 
 to see how  
these models can be tested by experiments. 

Roughly speaking, the tests can be classified into three broad categories:
modification of Newton's law; signatures in colliders; cosmology.
As usual in high energy physics, if the scale characterizing new physics
is too high  
then it cannot be  reached directly in collider experiments. In this case
cosmology is the only place where the effects of  new physics can 
be, indirectly, observed. 

As illustrated by numerous works in the last few years, 
Randall-Sundrum type cosmology is a very rich playground to study the very
peculiar consequences of the braneworld idea 
 in cosmology. As recalled  in this
contribution, its essential new features are:  a $\rho^2$ term 
in the generalized Friedmann equation, which dominates at  
high energies; ``dark radiation'',  {\it produced} during the high 
energy phase, and of 
potential relevance for observations, via the nucleosynthesis constraints
on the number of extra relativistic degrees of freedom.


\end{document}